\documentclass[prd,aps,preprint,nofootinbib,superscriptaddress]{revtex4}
\usepackage{epsfig}
\usepackage{amsmath}

\begin{document}

\preprint{BNL-NT-07/9} \preprint{RBRC-648}

\title{Single-Transverse Spin Asymmetry in Dijet Correlations
at Hadron Colliders}

\author{C.J. Bomhof}
\email{cbomhof@nat.vu.nl} \affiliation{
Department of Physics and Astronomy, Vrije Universiteit Amsterdam,\\
NL-1081 HV Amsterdam, the Netherlands}

\author{P.J. Mulders}
\email{mulders@few.vu.nl} \affiliation{
Department of Physics and Astronomy, Vrije Universiteit Amsterdam,\\
NL-1081 HV Amsterdam, the Netherlands}

\author{W. Vogelsang}
\email{vogelsan@quark.phy.bnl.gov}
\affiliation{Physics Department, Brookhaven National Laboratory,
Upton, NY 11973}
\author{F. Yuan}
\email{fyuan@quark.phy.bnl.gov}
\affiliation{RIKEN BNL Research Center, Building 510A, Brookhaven
National Laboratory, Upton, NY 11973}
\begin{abstract}
We present a phenomenological study of the single-transverse spin
asymmetry in azimuthal correlations of two jets produced nearly
``back-to-back'' in $pp$ collisions at RHIC. We properly take into
account the initial- and final-state interactions of partons that 
can generate this asymmetry in QCD hard-scattering.
Using distribution functions fitted to the existing single-spin data, 
we make predictions for various weighted single-spin asymmetries in 
dijet correlations that are now readily testable at RHIC.
\end{abstract}
\maketitle

\newcommand{\be}{\begin{equation}}
\newcommand{\ee}{\end{equation}}
\newcommand{\ben}{\[}
\newcommand{\een}{\]}
\newcommand{\beqn}{\begin{eqnarray}}
\newcommand{\eeqn}{\end{eqnarray}}
\newcommand{\Tr}{{\rm Tr} }

\section{Introduction}
Single-transverse spin asymmetries (SSAs) play a fundamental role
for our understanding of QCD in high-energy hadronic scattering.
They may be obtained for reactions in, for example, lepton-proton
or proton-proton scattering with one transversely polarized initial
proton, by dividing the difference of the cross sections for
the two settings of the transverse polarization by their
sum. There have been extensive experimental investigations of
such asymmetries~\cite{E704,star,balewski,phenix,brahms,hermes,dis}.
These have initiated much theoretical
progress, in particular within the last few years.

A particular focus has been on a class of single-spin observables
that are characterized by a large momentum scale $Q$ (for example,
the virtuality of the photon in deeply-inelastic scattering (DIS))
and by a much smaller, but also measured, transverse momentum
$q_\perp$. In such a ``two-scale'' situation, single-spin
asymmetries may arise at leading power, that is, not suppressed by
an inverse power of $Q$. For some of these cases, factorization
theorems have been established~\cite{ColSop81,JiMaYu04,ColMet04}
that allow to write the spin-dependent cross sections in terms of
parton distribution functions and/or fragmentation functions,
perturbative hard-scattering functions, and so-called soft
factors. A crucial feature is that the distribution functions and
the soft factor in this factorization are not integrated over the
transverse momenta of partons, because these in fact generate the
observed transverse momentum $q_\perp$. Among other things, the
observables may therefore provide valuable insights into the
dependence of parton distributions in nucleons on transverse
momentum. This becomes particularly interesting when the nucleon
is transversely polarized, because there may be correlations
between the nucleon spin vector, its momentum, and the parton's
transverse momentum. One particular correlation, known as the
``Sivers effect'' and described by so-called ``Sivers
functions''~\cite{Siv90}, is now widely believed to be involved in
a variety of observed hadronic single-spin phenomena.

Closer theoretical studies have revealed that the Sivers effect
plays an important role in QCD, beyond giving rise to phenomenological
functions to be used in the description of single-spin
asymmetries. A particularly interesting feature is that the Sivers effect is 
not universal in the usual sense, that is, it is not represented by 
universal probability functions convoluted with partonic hard-scattering 
cross sections. This might at first sight appear to
make the study of these functions less interesting. However, the
non-universality has in fact a clear physical origin, and its
closer investigation has turned out to be an extremely important
and productive development in QCD. In a nutshell, in order not to
be forced to vanish because of the time-reversal symmetry of QCD,
single-spin asymmetries require the presence of a
strong-interaction phase. For the Sivers functions this phase
originates from the ``gauge links'' in their
definition~\cite{Col02,BelJiYua02,BoeMulPij03}, which are
path-ordered exponentials of the gluon field that make the
functions gauge-invariant. In DIS, the gauge link may be viewed as
a rescattering of the parton in the color field of the nucleon
remnant. That such a final-state rescattering may generate the
phase required for a nonzero SSA in semi-inclusive hadron
production in DIS (SIDIS) was first discovered within a model
calculation~\cite{BroHwaSch02}.

Depending on the hard-scattering process, the ``rescattering''
will however manifest itself in different ways. For example, for
Drell-Yan lepton pair production in hadronic scattering, {\it
initial-state}, rather than final-state, interactions are
relevant. As a result, the phase provided by the gauge links is
opposite, and the Sivers functions for the Drell-Yan process have
opposite sign~\cite{BelJiYua02,BoeMulPij03,Col02,BroHwaSch02}.
In more general terms, the nontrivial ``universality" property of
the Sivers functions is the direct consequence of gauge
interactions in Quantum
Chromodynamics~\cite{Col02,BelJiYua02,BoeMulPij03}, and of the QCD
factorization theorems applying to the relevant hard
processes~\cite{ColSop81,JiMaYu04,ColMet04}. It is a remarkable
and fundamental QCD prediction that really tests all concepts we
know of for analyzing hard-scattering reactions in strong
interactions, and it awaits experimental verification.

While measurements of SSAs in SIDIS are now maturing and have
established the presence of Sivers-type contributions~\cite{hermes},
it will still be a while until precise single-spin measurements in the
relatively rare Drell-Yan process will become feasible at RHIC~\cite{rhic}
or elsewhere~\cite{dypax1}.
However, there are of course other hard-scattering reactions
in hadronic collisions for which single-transverse spin asymmetries
may be defined, and that may potentially be used in lieu of the
Drell-Yan process for testing the nontrivial ``universality" properties
of the Sivers functions. In \cite{BoeVog03}, it was proposed to use
the SSA in azimuthal correlations of two jets produced in $pp$ collisions
at RHIC to learn about the Sivers functions. To a first approximation,
such dijets are produced by $2\to 2$ partonic QCD hard-scattering.
With collinear kinematics, the jets are exactly ``back-to-back'' in
the plane perpendicular to the initial beam directions and thus
separated by $180^{\rm o}$ in azimuthal angle in this plane. Partonic
transverse momenta will generate deviations from this topology,
because they will lead on average to a non-vanishing net transverse
momentum $q_\perp$ of the jet pair, much smaller than each of the jet
transverse momenta $P_\perp$ individually. The observable is therefore
of the ``two-scale'' type described above, and as was shown
in \cite{BoeVog03}, if one proton is polarized, a single-spin
asymmetry may be defined that is in principle sensitive to the
Sivers functions. As a caveat, factorization of the spin-dependent
cross section for this observable in terms of transverse-momentum-dependent
(TMD) functions still remains to be established.

Unlike the relatively simple cases of SIDIS and the Drell-Yan process,
where either final-state or initial-state interactions contribute to the
Sivers asymmetry, both are present for dijet production \cite{Sterman}.
This complicates the analysis of the process-dependence of the Sivers
functions considerably, but at the same time it also makes it much
more interesting from a theoretical point of view, because the
interactions in this case are ``truly QCD'', that is, they involve
the detailed gauge structure of the theory, including for example
its non-abelian nature. At the time of \cite{BoeVog03}, the
process-dependence had not yet been worked out for the case of the SSA in
dijet production, so that phenomenological predictions had to remain
qualitative, at best. Over the last year, however, there has been
extensive work on deriving and clarifying the structure of the gauge links
for this and related processes in $pp$ 
collisions \cite{mulders,bomhof,bacchetta}.
Indeed, the resulting structure is far more complicated than that in
SIDIS or the Drell-Yan process. However, it turns out that if one
takes a certain weighted integral (``moment'') of the asymmetry,
remarkable simplifications occur. This moment is defined by integrating
the spin-dependent cross section with a factor $\sin\delta$, where
$\delta$ is the azimuthal imbalance between the two jets ($\delta =0$
if the jets are exactly back-to-back in azimuth). For each of the various
contributing $2\to 2$ partonic channels, the gauge link then
essentially collapses into a set of simple color factors that
multiply contributions from color-gauge invariant subamplitudes
to the given partonic process. One may, in fact, for convenience choose
to absorb these factors into the hard-scattering functions, and {\it define}
the Sivers functions as the functions measured in the SSA in SIDIS. In
this way, the net effect of the gauge links on the $\sin\delta$-moment of
the spin-dependent cross section is to generate new partonic hard-scattering
functions that are different from the usual spin-independent ones, but
that are actually similarly simple in structure.

At the same time, taking the moment of the factorized spin-dependent
cross section leads to a new expression that involves only a certain
moment of the Sivers functions in partonic transverse momentum $k_\perp$,
rather than the fully transverse-momentum dependent functions themselves.
As was shown in~\cite{BoeMulPij03}, these
$k_\perp$-moments of the Sivers functions are directly related to
twist-three quark-gluon correlation functions first introduced
in~\cite{Efremov,qiu} to describe the SSA for single-inclusive hadron
production in hadronic scattering. By now, quite some knowledge about
these correlation functions has been gathered from phenomenological
studies~\cite{twist3-06} of the corresponding data.

The upshot of all this is that it has now become possible for the first time
to make predictions for the $\sin\delta$-moment of the single-transverse
spin asymmetry in dijet production at RHIC that correctly take into
account the process-dependence of the Sivers functions and incorporate
phenomenological information on some properties of the functions that is
available from other measurements. This is the goal of this note.

\section{Spin-dependent cross section and $\sin\delta$-moment}
We study azimuthal correlations of two jets produced nearly
``back-to-back'' in a hadronic collision.
More specifically, we are interested in situations in which the
sum of the two jet transverse momenta,
$\vec{q}_\perp\equiv\vec{P}_{1\perp}+\vec{P}_{2\perp}$ (or a
component or projection thereof), is measured and small, 
while both individual jet transverse momenta are large and similar.
We will therefore approximate $\vec P_{1\perp} = -\vec P_{2\perp}\equiv 
\vec P_\perp$ wherever possible. For the lengths of these momentum vectors 
we will simply write $P_\perp = \vert \vec P_\perp\vert$ and $q_\perp = 
\vert \vec q_\perp\vert$.
The differential cross section for the process with one transversely
polarized hadron contains terms of the form
\begin{equation}
\frac{2\pi\,d^6\sigma}{d\eta_1\,d\eta_2\,dP_\perp^2\,d\phi_1\,d^2q_\perp}
=\frac{2\pi\,d^6\sigma_{UU}}{d\eta_1\,d\eta_2\,dP_\perp^2\,d\phi_1\,d^2q_\perp}
+\hat e_z\cdot\left(\vec S_\perp\times\hat q_\perp\right)
\frac{2\pi\,d^6\sigma_{TU}}{d\eta_1\,d\eta_2\,dP_\perp^2\,d\phi_1\,
d^2q_\perp} \; ,
\label{sigma}
\end{equation}
where $\hat e_z$ is a unit vector in the direction of the polarized
proton, $\vec S_\perp$ is the transverse spin vector of the
polarized proton, and $\eta_1$ and $\eta_2$ are the pseudo-rapidities of
the two jets. The first term in Eq.~(\ref{sigma}) represents the 
unpolarized (or spin-averaged) cross section, while the second term is the
single-transverse-spin dependent one. We note that the angular
dependence of the spin-dependent term is $\vert \vec S_\perp\vert
\,\sin(\phi_b - \phi_S)$, where $\phi_b = (\phi_1+\phi_2)/2$ is the 
so-called bisector-angle of the two jets, which (approximately)
corresponds to the direction of $\vec q_\perp$. For this reason one 
may also choose to integrate the six-fold differential cross section in 
Eq.~(\ref{sigma}) over $\phi_1$, keeping $\vec q_\perp$ fixed. 

As a generalization of the SIDIS and Drell-Yan cases
\cite{JiMaYu04}, we can write down a factorization formula for the
differential cross section at small imbalance ($\vec q_\perp \ne 0$)
of the jets, in
terms of TMD parton distributions, soft factors, and hard-scattering
functions~\cite{VogYua05}. We remind the
reader that such a factorization still remains to be proven. In
this paper, we will not discuss the details of factorization
issues related to the dijet correlations.
As we mentioned above, significant simplifications occur when the
imbalance of the two jets is integrated out by taking certain
moments. For example, integrating the spin-independent
differential cross section over all $\vec q_\perp$, its expression
reverts to the standard collinear factorization formula for dijet
production,
\begin{eqnarray}
\langle 1\rangle_{UU} &\equiv &\int d^2\vec{q}_\perp
\frac{2\pi\,d^6\sigma_{UU}}
{d\eta_1\,d\eta_2\,dP_\perp^2\,d\phi_1\,d^2\vec{q}_\perp}=
\frac{2\pi\,d^4\sigma_{UU}}{d\eta_1\,d\eta_2\,dP_\perp^2\,d\phi_1}
\nonumber \\
& = & \sum\limits_{ab}
x_af_a(x_a)x_bf_b(x_b)H_{ab\to cd}^{uu}(\hat{s},\hat{t},\hat{u}) \ ,
\label{jeteq}
\end{eqnarray}
where the $f_i$ denote the usual collinear (light-cone) parton
distribution functions for parton type $i=q,\bar{q},g$. We have
assumed here that these are recovered by integration of the
corresponding TMD distribution functions over all partonic
transverse momentum, and we disregard the renormalization
properties of the operators defining these distributions and the
scale dependence introduced by renormalization.
However, all these effects can be systematically included
accordingly.

The factors $H_{ab\to cd}^{uu}$ in Eq.~(\ref{jeteq}) are the customary
hard-scattering cross sections $d\hat\sigma^{ab\to cd}/d\hat t$ for
the partonic processes $ab\to cd$ (for a compilation, see, for example,
Ref.~\cite{bomhof}).
They are functions of the partonic
Mandelstam variables, $\hat{s}=(p_a+p_b)^2$, $\hat{t}=(p_a-p_c)^2$,
$\hat{u}=(p_a-p_d)^2$, with obvious notation of the parton momenta.
In terms of the hadronic center-of-mass energy $\sqrt{s}$ and the jet
transverse momenta and pseudo-rapidities, one has
$\hat s=x_ax_b s$, $\hat t=-P_\perp^2\left(e^{\eta_2-\eta_1}+1\right)$, $\hat
u=-P_\perp^2\left(e^{\eta_1-\eta_2}+1\right)$, where the partonic momentum
fractions are fixed by $x_a=\frac{P_\perp}{\sqrt{s}}
\left(e^{\eta_1}+e^{\eta_2}\right)$, $x_b=\frac{P_\perp}{\sqrt{s}}
\left(e^{-\eta_1}+e^{-\eta_2}\right)$.

Next we turn to the single-transverse-spin dependent differential cross
section $d\sigma_{TU}$ in Eq.~(\ref{sigma}), which can be further
simplified by taking a moment in 
\begin{equation}
P_\perp\sin\delta 
= \frac{\hat e_z\cdot (\vec P_\perp\times \vec q_\perp)}{P_\perp} \; ,
\label{weight}
\end{equation}
where $\delta = \pi - (\phi_2-\phi_1)$ measures how far the two jets are
away from the back-to-back configuration. Within our approximations, 
the weight $P_\perp\,\sin\delta$ corresponds to a weight in $q_\perp$. 
One finds~\cite{bacchetta}:
\begin{eqnarray}
\label{qtmp} 
\left\langle \frac{2\,P_\perp\sin\delta}{M_P}\right\rangle_{TU} &\equiv& 
\vert \vec S_\perp\vert 
\int d^2\vec{q}_\perp\ \frac{q_\perp}{M_P}\,\frac{2\pi\,d^6\sigma_{TU}}
{d\eta_1\,d\eta_2\,dP_\perp^2\,d\phi_1\,d^2\vec{q}_\perp}
\nonumber \\[2mm]
&=& \vert \vec S_\perp\vert 
\sum_{ab} x_a\,\frac{-1}{M_P}\,gT_{F\,a}(x_a)\,x_bf_b(x_b)
\,H_{ab\to cd}^{\rm sivers}(\hat{s},\hat{t},\hat{u})\ ,
\end{eqnarray}
where $M_P$ is the proton mass, $g$ is the strong coupling constant 
and the $T_{F\,a}(x)$ are
the Qiu-Sterman matrix elements or quark-gluon correlation
functions \cite{qiu}, defined as
\begin{eqnarray}
\label{QSme}
T_{F\,a}(x) = \int\frac{d\xi^-d\eta^-}{4\pi}e^{i(xP^+\eta^-)}
\epsilon_\perp^{\beta\alpha}S_{\perp\beta} 
\left\langle PS|\overline\psi{}^a(0)\gamma^+ F_{\alpha}^{\ +}(\xi^-)
\psi^a(\eta^-)|PS\right\rangle \; ,
\end{eqnarray}
with the quark fields $\psi^a$ (for flavor $a$) and the gluon field
strength tensor $F_{\alpha}^{\ +}$. The $T_{F\,a}(x)$ matrix elements
enter because they are related to the $k_\perp$-moment of the (SIDIS)
Sivers function for quark flavor $a$ \cite{BoeMulPij03}, that is,
$g\,T_{F\,a}(x) = -2M\,f_{1T,a}^{\perp (1)}(x)$. We note that
there could also be contributions by purely gluonic ``$ggg$'' correlation 
functions. These will be ignored for now, so that the label $a$ in 
Eq.~(\ref{qtmp}) runs only over quarks and anti-quarks. Furthermore, 
there is actually a second contribution to the single-transverse-spin
dependent cross section, which involves the scattering of transversely 
polarized quarks from both the polarized proton (transversity distribution, 
$\delta f$ or $h_1$) and from the unpolarized proton 
(Boer-Mulders function $h_1^\perp$~\cite{boermulders}). The latter functions 
are also effects of initial- and final-state interactions and appear in 
a matrix element for unpolarized protons similar to Eq.~\eqref{QSme}. 
Like the Sivers-type ``$ggg$'' correlation functions, we will also ignore
the contributions associated with the Boer-Mulders functions 
in the present study, even though a future more 
detailed analysis of forthcoming experimental data may well require 
to take all of these into account.

The relevant hard-scattering functions $H_{ab\to cd}^{\rm sivers}$
have been calculated in~\cite{bacchetta,bomhof}, where they were termed
``gluonic-pole cross sections'' due to their association with the
Qiu-Sterman or gluonic pole matrix elements~\cite{qiu}. We have also 
independently reproduced~\cite{qvyprep} 
the $H_{ab\to cd}^{\rm sivers}$ within a model
calculation along the lines of Ref.~\cite{BroHwaSch02}. For convenience,
we list the ones that will be relevant for our numerical calculations
presented below:
\begin{align}
\label{hsivers} H_{qq'\to qq'}^{\rm sivers}(\hat s,\hat t,\hat u)
&=\frac{\alpha_s^2\pi}{\hat
s^2}\frac{N_c^2-5}{4N_c^2}\;\frac{2(\hat s^2+\hat u^2)}{\hat
t^2}\ ,\nonumber\\
H_{q\bar q'\to q\bar q'}^{\rm sivers}(\hat s,\hat t,\hat u)
&=\frac{\alpha_s^2\pi}{\hat
s^2}\left(-\frac{N_c^2-3}{4N_c^2}\right)\frac{2(\hat s^2+\hat
u^2)}{\hat t^2}\ ,\nonumber\displaybreak[0]\\
H_{q\bar q\to q'\bar q'}^{\rm sivers}(\hat s,\hat t,\hat u)
&=\frac{\alpha_s^2\pi}{\hat
s^2}\frac{N_c^2+1}{4N_c^2}\frac{2(\hat t^2+\hat
u^2)}{\hat s^2} \ ,\nonumber\displaybreak[0]\\
H_{qq\to qq}^{\rm sivers}(\hat s,\hat t,\hat u)
&=\frac{\alpha_s^2\pi}{\hat s^2}
\left\{\frac{N_c^2-5}{4N_c^2}\left[\frac{2(\hat s^2+\hat u^2)}{\hat t^2}
+\frac{2(\hat s^2+\hat t^2)}{\hat u^2}\right]
+\frac{N_c^2+3}{4N_c^3}\frac{4\hat s^2}{\hat t\hat u}\right\} \ ,
\nonumber\displaybreak[0]\\
H_{q\bar q\to q\bar q}^{\rm sivers}(\hat s,\hat t,\hat u)
&=\frac{\alpha_s^2\pi}{\hat
s^2}\left\{-\frac{N_c^2-3}{4N_c^2}\frac{2(\hat s^2+\hat u^2)}{\hat t^2}
+\frac{N_c^2+1}{4N_c^2}\frac{2(\hat u^2+\hat t^2)}{\hat s^2}
-\frac{N_c^2+1}{4N_c^3}\frac{4\hat u^2}{\hat s\hat t}\right\} \ ,
\nonumber\displaybreak[0]\\
H_{qg\to qg}^{\rm sivers}(\hat s,\hat t,\hat u)
&=\frac{\alpha_s^2\pi}{\hat
s^2}\left\{-\frac{N_c^2}{4(N_c^2-1)}\frac{2(\hat s^2+\hat
u^2)}{\hat t^2}\left[\frac{\hat s}{\hat u}-\frac{\hat u}{\hat
s}\right]-\frac{1}{2(N_c^2-1)}\frac{2(\hat s^2+\hat u^2)}{\hat
t^2}\right.\nonumber\\
&~~~~~~~~~~~~\left.-\frac{1}{4N_c^2(N_c^2-1)}\frac{2(\hat s^2+\hat
u^2)}{\hat
s\hat u}\right\} \ ,\nonumber\\
H_{q\bar q\to gg}^{\rm sivers}(\hat s,\hat t,\hat u)
&=\frac{\alpha_s^2\pi}{\hat
s^2}\left\{-\frac{1}{2N_c}\frac{2(\hat t^2+\hat u^2)}{\hat
s^2}+\frac{N_c}{4}\frac{2(\hat t^2+\hat u^2)}{\hat
s^2}\left[\frac{\hat t}{\hat u}+\frac{\hat u}{\hat t}\right]\right.\nonumber\\
&~~~~~~~~~~~~\left.-\frac{2N_c^2+1}{4N_c^3}\frac{2(\hat t^2+\hat
u^2)}{\hat t\hat u}\right\} \ ,
\end{align}
where $N_c=3$ is the number of colors. Hard-scattering functions
corresponding to gluonic correlation functions were also
calculated in~\cite{bomhof}, but are not taken into account in our
present study as we discussed above. Comparing the above functions 
with the usual spin-averaged hard-scattering
functions~\cite{bomhof}, one can see that they essentially differ
in the color factors that multiply terms with similar kinematic
structure. This is the net result of the
combined effects from the initial and final state interactions in
dijet production in hadronic reactions.

\section{Phenomenological results for RHIC}
\begin{figure}[t]
\epsfig{figure=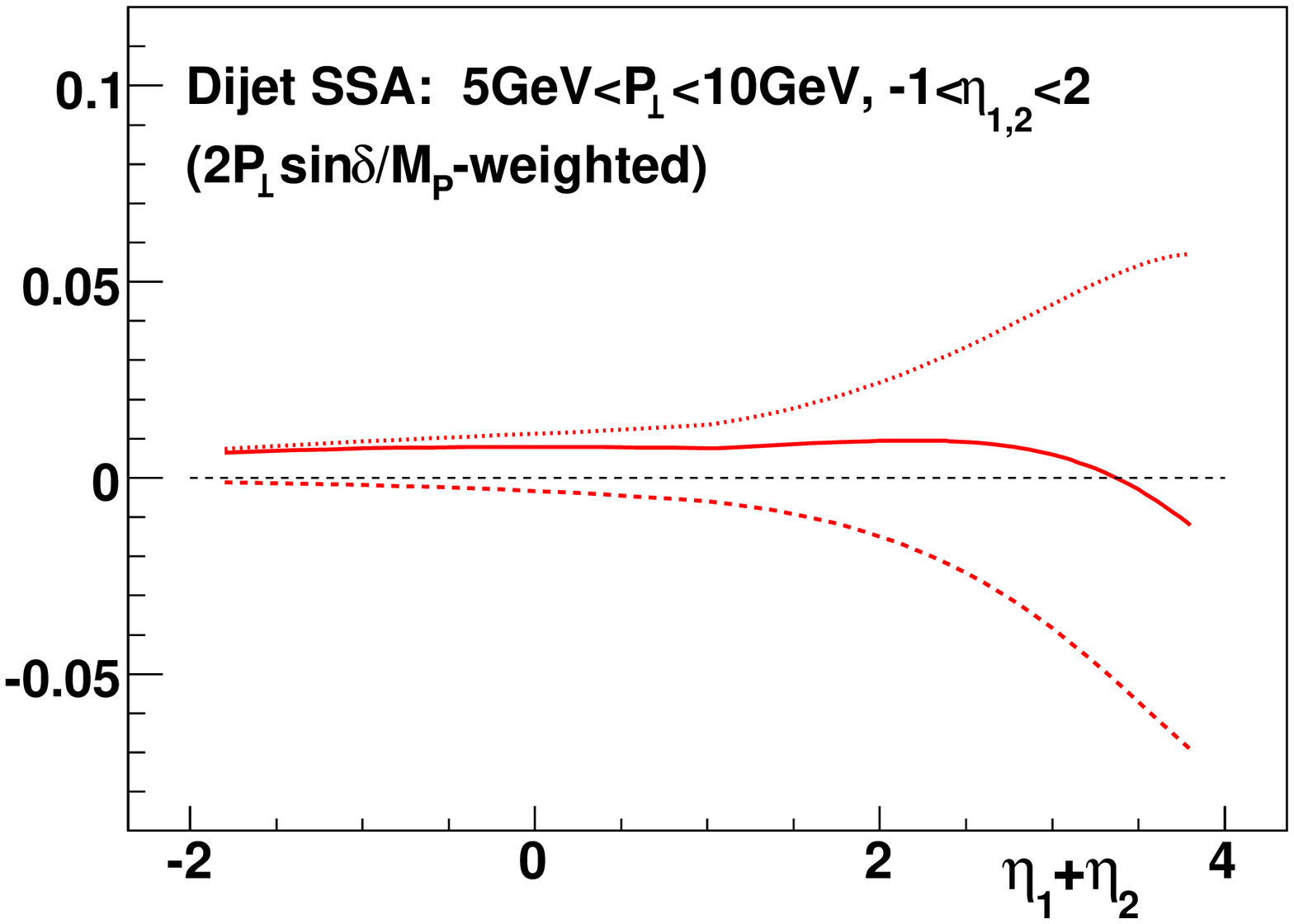,width=8cm, height=7cm}
\epsfig{figure=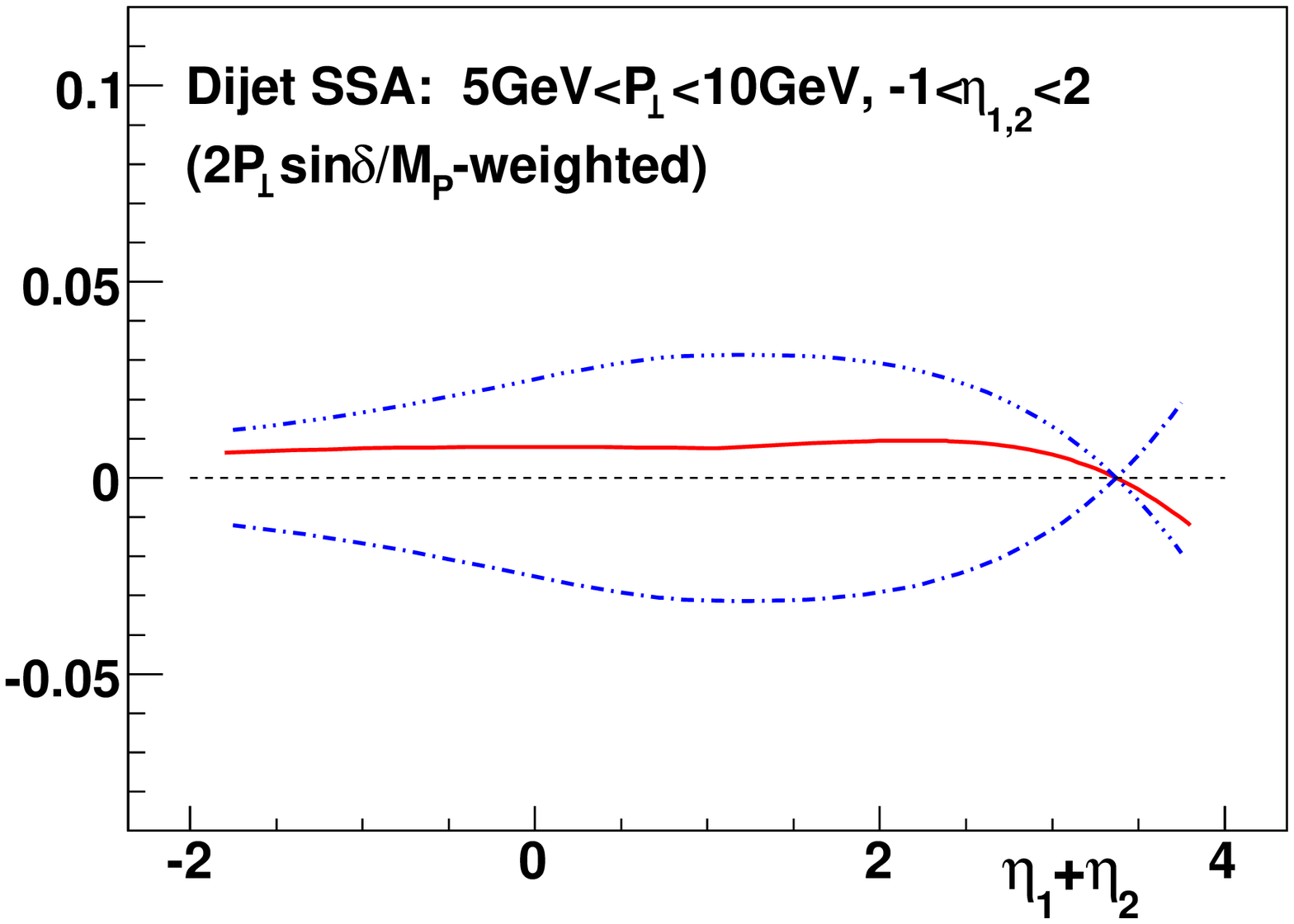,width=8cm, height=7cm}\vskip -0.4cm
\caption{\it The weighted single-transverse-spin asymmetry for
dijet-correlations in polarized proton-proton scattering at RHIC,
as a function of the sum of the jet pseudo-rapidities. The jet
transverse momenta are in the range $5-10$~GeV. The solid line is
our main result, based on Eqs.~(\ref{hsivers})-(\ref{tfdist}). In
the left panel we also plot the individual contributions due to the
$u$-quark (dashed line) and $d$-quark (dotted line) Qiu-Sterman
matrix elements. For comparison, in the
right panel we also show the results that would be obtained if the
Sivers functions contributing to dijet production were the same as
those in SIDIS (upper curve) or the Drell-Yan process (lower
curve). \label{fig1}}
\end{figure}

We now use the above formulas to obtain some predictions for the SSA in
dijet-production at RHIC. A SSA for our
$(2P_\perp\sin\delta/M_P)$-moment (i.e. $q_\perp/M_P$-moment)
can be defined as (from now on we choose $\vert \vec S_\perp\vert=1$)
\begin{eqnarray}
\frac{\langle 2P_\perp\sin\delta/M_P\rangle_{TU}}{\langle
1\rangle_{UU}}=\frac{\sum_{ab} x_a
\frac{-1}{M_P}gT_{F\,a}(x_a)x_bf_b(x_b)H_{ab\to cd}^{\rm
sivers}(\hat{s},\hat{t},\hat{u})} {\sum_{ab} x_a
f_a(x_a)x_bf_b(x_b) H_{ab\to cd}^{uu}(\hat{s},\hat{t},\hat{u})} \ .
\end{eqnarray}
The quark-gluon correlation functions $T_{F\,a}$ have
recently been fitted~\cite{twist3-06} to data~\cite{E704,star,brahms}
for the SSA in single-inclusive hadron production in hadronic collisions.
Two sets of $T_{F\,a}$ were presented in~\cite{twist3-06}. For definiteness,
we choose set~I, which is a two-flavor fit with valence $u$ and
$d$-quark $T_F$ distributions only. For these the following
parameterizations were obtained in~\cite{twist3-06}:
\begin{equation}
\label{tfdist} T_{F\,u_v}(x)=0.275 x^{0.508} (1-x)^{0.399} {u_v}(x)
\; , \;\;\; T_{F\,d_v}(x)=-0.365 x^{-0.108} (1-x)^{0.287} {d_v}(x)
\ ,
\end{equation}
where the ${u_v},{d_v}$ are the corresponding unpolarized valence
quark distributions. For the latter, as for all other unpolarized
parton distributions we need, we choose the CTEQ5L
parameterizations \cite{cteq5l}. We plot the resulting weighted
asymmetry in Fig.~\ref{fig1}. The kinematics are chosen to
correspond to current measurements at STAR: both jets have
transverse momenta $5\,{\rm GeV}\leq P_\perp \leq 10\,{\rm GeV}$
and pseudo-rapidities $-1\leq \eta_i \leq 2$. The asymmetry is
plotted as a function of the sum of the two jet pseudo-rapidities.
We have chosen the hard scale $\mu=P_\perp$ in the strong coupling
constant and the unpolarized parton distributions.

For comparison, we also show in Fig.~\ref{fig1} the asymmetries
that one would have expected if the Sivers functions relevant for
dijet production were identical to the ones in SIDIS or the
Drell-Yan process. In the framework of our present calculation,
the corresponding partonic hard-scattering functions would in this
case be identical to the spin-averaged functions $H_{ab\to
cd}^{uu}$, or to their negatives. The dotted-dashed lines in the
right panel of Fig.~\ref{fig1} represent these cases. They
essentially correspond to the earlier predictions
of~\cite{BoeVog03} and~\cite{VogYua05} that were made when the
process-dependence of the Sivers functions had not yet been worked
out for the dijet case. One can see that when the correct
process-dependence is incorporated, the asymmetry overall becomes
smaller by about a factor two, which can be traced back to the
color factors for the dominant subprocess $qg\to qg$. The sign is
identical to the case when the Sivers functions for dijet
production are assumed to be ``DIS-like'', implying that in a
sense final-state interactions dominate over initial-state ones.

In principle, one might verify experimentally the
process-dependence of the Sivers functions by discriminating
between the various curves in Fig.~\ref{fig1} and confirming the
QCD-predicted result shown by the solid line. In practice, this
may require good knowledge of the $T_F$ distributions, and an
understanding of issues like scale evolution and higher-order
corrections. A closer analysis reveals that the asymmetry in
Fig.~\ref{fig1} is the result of rather significant cancellations
between contributions of opposite signs by $u$ and $d$ quarks. To
show this, we display their individual contributions separately in
the left panel of Fig.~\ref{fig1}.

We finally also briefly discuss a related type of SSA
in dijet production at RHIC. In \cite{VogYua05}, a differently
weighted SSA was considered, defined in the following way:
\begin{equation}
\label{sign}
\langle{\rm Sgn}(\delta)\rangle_{TU}
\equiv \int d^2\vec{q}_\perp\, {\rm Sgn}(\delta)
\frac{d^5\sigma_{TU}}
{d\eta_1d\eta_2dP_\perp^2d^2\vec{q}_\perp}  \ ,
\end{equation}
where ${\rm Sgn}(x)$ is the sign function. This asymmetry has the property
that the weight only depends on the azimuthal separation of the two
jets, but not on their transverse momenta. This may be an advantage
for experimental measurements when the jet energy scale is not precisely
known. When applied to the TMD factorized expression for the spin-dependent
cross section, the moment defined in Eq.~(\ref{sign}) leads to an
expression different from~(\ref{qtmp}). One finds in fact that in general the
transverse-momentum dependences of the various functions do not
completely decouple anymore, so that the final result can in general
no longer be expressed in terms of only functions of light-cone momentum
fractions. If one assumes for simplicity~\cite{VogYua05}, however, that
the only relevant dependence on transverse momentum resides in the Sivers
functions, the resulting 
expression again resembles a collinearly-factorized one:
\begin{equation}
\frac{\langle{\rm Sgn}(\delta)\rangle_{TU}}{\langle
1\rangle_{UU}}=\frac{\sum_{ab} x_a q_{Ta}^{\rm
(1/2)}(x_a)x_bf_b(x_b)H_{ab\to cd}^{\rm sivers}
(\hat{s},\hat{t},\hat{u})}{\sum_{ab}
x_a f_a(x_a)x_bf_b(x_b)H_{ab\to cd}^{uu}(\hat{s},\hat{t},\hat{u})} \ ,
\label{eqan}
\end{equation}
where the $q_{Ta}^{\rm (1/2)}$ are the so-called ``$1/2$-moments'' of the
Sivers functions for SIDIS and are defined as
\begin{eqnarray}\label{eq7}
q_{Ta}^{(1/2)}(x)&\equiv&\int d^2k_\perp\frac{|\vec{k}_\perp|}{M}
q_{Ta}^{\rm SIDIS}\left({x}, k_{\perp}\right) \ .
\end{eqnarray}
In the above equations, we have used the notation ``$q_{Ta}$'' for the
Sivers function for quark flavor $a$; its definition is identical to that in
\cite{BoeMulPij03}: $q_{Ta}\equiv f_{1T,a}^{\perp}$.
In~\cite{VogYua05}, the hard-scattering functions were chosen to
be the same as the spin-averaged ones, with opposite sign. In
light of the above discussions, one now would like to update the
predictions for the asymmetry in~(\ref{eqan}). 
We will use the $H_{ab\to cd}^{\rm sivers}$ given in
Eqs.~(\ref{hsivers}). We note that it remains to be established
that the same $H_{ab\to cd}^{\rm sivers}$ do contribute in this
case. This is not a priori clear, because the general gauge link
structure is very complex in the general TMD case, and it has so
far only been demonstrated that the $H_{ab\to cd}^{\rm sivers}$
apply when the $\sin\delta$ moment is taken. For now,
we just conjecture that use of the $H_{ab\to cd}^{\rm sivers}$
of~\cite{bacchetta,bomhof} is justified in this case; a closer
discussion of this issue is left for future work~\cite{qvyprep}.

\begin{figure}[t]
\epsfig{figure=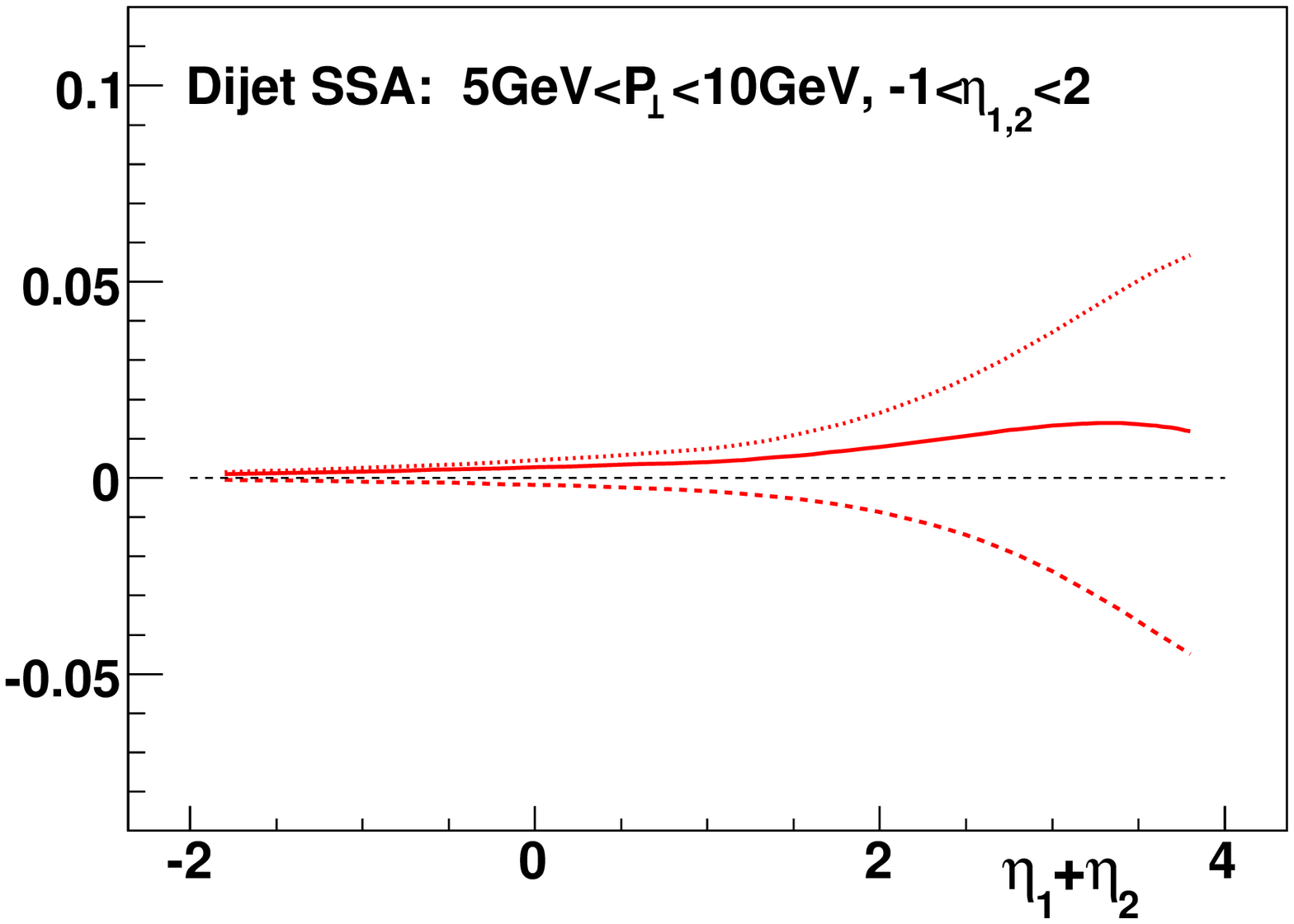,width=8cm, height=7cm}
\epsfig{figure=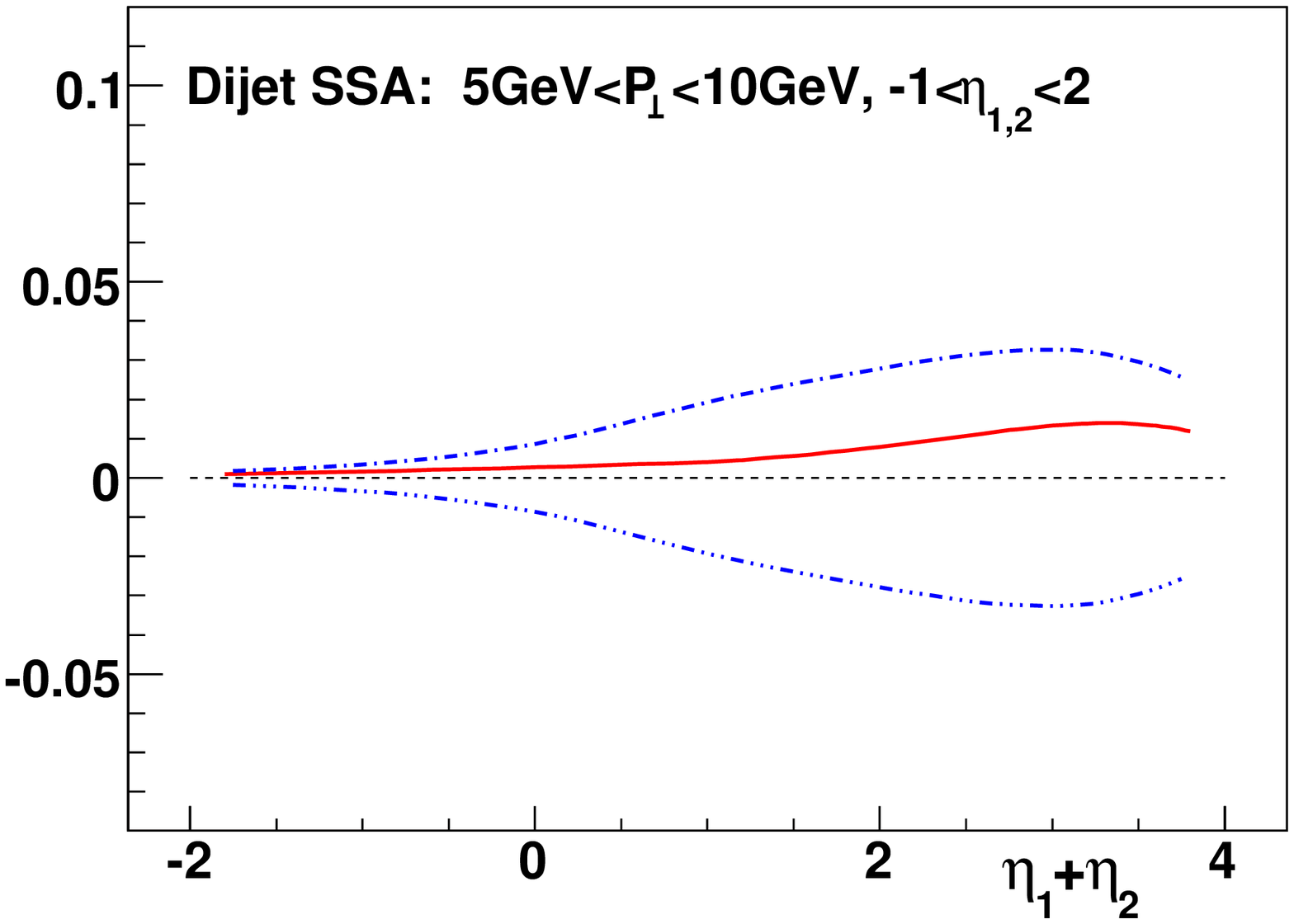,width=8cm, height=7cm}\vskip -0.4cm
\caption{\it Same as Fig.~1, but for the SSA defined in
Eq.~(\ref{eqan}). \label{fig2}}
\end{figure}

The 1/2-moments of the quark Sivers functions were determined
in~\cite{VogYua05} by a fit to the HERMES data~\cite{hermes}
for the Sivers-type single-spin asymmetry:
\begin{equation}
u_{T}^{\rm (1/2)}(x)=-0.75 x(1-x) u(x),~~~d_{T}^{\rm
(1/2)}(x)=2.76x(1-x)d(x) \ .
\end{equation}
These results correspond to set~II presented in~\cite{VogYua05}.
In Fig.~\ref{fig2}, we show predictions for the asymmetry for the
dijet-correlation defined in~(\ref{eqan}) at RHIC, based on this
parameterization. We find that the asymmetry shares many features
with the one shown for the $(2P_\perp\sin\delta/M_P)$-moment in
Fig.~\ref{fig1}. We note that first preliminary experimental
data for this asymmetry have now been presented by
STAR~\cite{balewski}, which are so far consistent with a vanishing
asymmetry.

\section{Conclusions} 
In summary, we have studied
single-transverse spin asymmetries in dijet-correlations at RHIC,
making use of the recently derived partonic hard-scattering cross
sections that properly incorporate the initial- and final-state
interactions, and of distribution functions fitted to existing
data for single-spin asymmetries. We have found that the initial-
and final-state interactions tend to decrease the magnitude of the
asymmetry at RHIC with respect to earlier estimates that assumed
the Sivers functions for this observable to be identical to the
Sivers functions in SIDIS or the Drell-Yan process. Overall, the
resulting asymmetries turn out to be more dominated by the
final-state interactions, and hence more ``DIS-like''.

With experimental data on dijet single-spin asymmetries now forthcoming,
it will be interesting to perform detailed comparisons
with the theoretical expectations. Other observables, such as
the SSAs in dihadron production $pp\to h_1 h_2 X$ or in photon-plus-jet
production $pp\to \gamma\,{\mathrm{jet}}\, X$, will also be
extremely interesting. It will be important to further develop
the theoretical framework for all these observables,
by addressing issues like TMD factorization, higher orders, soft factors,
and Sudakov suppression, in particular.

{\bf Acknowledgments.} We thank Jan Balewski and Steve Vigdor for
valuable communications about the recent STAR measurements and the
future prospects. We are also grateful to Daniel Boer and Jianwei
Qiu for discussions. W.V. and F.Y. are grateful to RIKEN,
Brookhaven National Laboratory and the U.S. Department of Energy
(contract number DE-AC02-98CH10886) for providing the facilities
essential for the completion of their work.
This work has benefited from the research program of the EU Integrated
Infrastructure Initiative Hadron Physics (RII3-CT-2004-506078). The work
of C.B. is supported by the Foundation for Fundamental
Research of Matter (FOM) and the National Organization for Scientific
Research (NWO).


\begin{thebibliography}
\frenchspacing

\bibitem{E704} for fixed-target SSA data, see for example:
D.~L.~Adams {\it et al.}  [E581 and E704 Collaborations],
  Phys.\ Lett.\ B {\bf 261}, 201 (1991);
 D.~L.~Adams {\it et al.}  [FNAL-E704 Collaboration],
  Phys.\ Lett.\ B {\bf 264}, 462 (1991);
K.~Krueger {\it et al.}, Phys.\ Lett.\ B {\bf 459}, 412 (1999).

\bibitem{star}
  J.~Adams {\it et al.}  [STAR Collaboration],
  Phys.\ Rev.\ Lett.\  {\bf 92}, 171801 (2004);
L.~Nogach, talk presented at the
``17th International Spin Physics Symposium (Spin 2006)'',
Kyoto, Japan, October 2-7, 2006, arXiv:hep-ex/0612030.

\bibitem{balewski} J.~Balewski, talk presented at the
``17th International Spin Physics Symposium (Spin 2006)'', Kyoto,
Japan, October 2-7, 2006, arXiv:hep-ex/0612036;
S. Vigdor, talk presented at the Annual
Meeting of the Division of Nuclear Physics of the American
Physical Society, Nashville, Tennessee, October 25-28, 2006.

\bibitem{phenix}
  S.~S.~Adler  [PHENIX Collaboration],
Phys.\ Rev.\ Lett.\  {\bf 95}, 202001 (2005);
M.~Chiu, talk presented at the
``17th International Spin Physics Symposium (Spin 2006)'',
Kyoto, Japan, October 2-7, 2006, arXiv:nucl-ex/0701031;
C.~A.~Aidala, Ph.D. Thesis, Columbia U.,
  arXiv:hep-ex/0601009.

\bibitem{brahms}
  F.~Videbaek  [BRAHMS Collaboration],
AIP Conf.\ Proc.\  {\bf 792}, 993 (2005);
J.~H.~Lee [BRAHMS Collaboration], talk presented at the
``17th International Spin Physics Symposium (Spin 2006)'',
Kyoto, Japan, October 2-7, 2006.

\bibitem{hermes}
A.~Airapetian {\it et al.}  [HERMES Collaboration],
Phys.\ Rev.\ Lett.\  {\bf 84}, 4047 (2000);
  Phys.\ Rev.\ Lett.\  {\bf 94}, 012002 (2005);
M. Diefenthaler [HERMES Collaboration],
talk presented at the
``17th International Spin Physics Symposium (Spin 2006)'',
Kyoto, Japan, October 2-7, 2006, arXiv:hep-ex/0612010.

\bibitem{dis}
A.~Bravar  [Spin Muon Collaboration],
Nucl.\ Phys.\ A {\bf 666}, 314 (2000);
H.~Avakian  [CLAS Collaboration], talk presented at the RBRC
workshop ``Single-Spin Asymmetries'', Brookhaven National
Laboratory, Upton, New York, June 1-3, 2005, RBRC proceedings
volume 75 (BNL-74717-2005);
  V.~Y.~Alexakhin {\it et al.}  [COMPASS Collaboration],
  Phys.\ Rev.\ Lett.\  {\bf 94}, 202002 (2005);
  E.~S.~Ageev {\it et al.}  [COMPASS Collaboration],
  arXiv:hep-ex/0610068;
F.~Bradamante [COMPASS Collaboration], talk presented at the
``17th International Spin Physics Symposium (Spin 2006)'',
Kyoto, Japan, October 2-7, 2006.

\bibitem{ColSop81}
J.~C.~Collins and D.~E.~Soper,
Nucl.\ Phys.\ B {\bf 193}, 381 (1981) [Erratum-ibid.\ B {\bf 213},
545 (1983)];
Nucl.\ Phys.\ B {\bf 197}, 446 (1982).


\bibitem{JiMaYu04}
  X.~Ji, J.~P.~Ma and F.~Yuan,
  Phys.\ Rev.\ D {\bf 71}, 034005 (2005);
Phys.\ Lett.\ B {\bf 597}, 299 (2004).


\bibitem{ColMet04}
J.~C.~Collins and A.~Metz,
Phys.\ Rev.\ Lett.\  {\bf 93}, 252001 (2004).

\bibitem{Siv90}
D.~W.~Sivers,
Phys.\ Rev.\ D {\bf 41}, 83 (1990);
Phys.\ Rev.\ D {\bf 43}, 261 (1991).


\bibitem{Col02}
J.~C.~Collins,
Phys.\ Lett.\ B {\bf 536}, 43 (2002).


\bibitem{BelJiYua02}
X.~Ji and F.~Yuan,
Phys.\ Lett.\ B {\bf 543}, 66 (2002);
A.~V.~Belitsky, X.~Ji and F.~Yuan,
Nucl.\ Phys.\ B {\bf 656}, 165 (2003).

\bibitem{BoeMulPij03}
D.~Boer, P.~J.~Mulders and F.~Pijlman,
Nucl.\ Phys.\ B {\bf 667}, 201 (2003).

\bibitem{BroHwaSch02}
S.~J.~Brodsky, D.~S.~Hwang and I.~Schmidt,
Phys.\ Lett.\ B {\bf 530}, 99 (2002);
Nucl.\ Phys.\ B {\bf 642}, 344 (2002).

\bibitem{rhic} C. Aidala  {\it et al.},
{\tt http://spin.riken.bnl.gov/rsc/report/masterspin.pdf},
{\it Research Plan for Spin
Physics at RHIC}.

\bibitem{dypax1} GSI-PAX Collab.,
P.~Lenisa and F.~Rathmann (spokespersons) {\em et al.}, Technical
Proposal, arXiv:hep-ex/0505054; Y.~Goto, talk presented at the
``Workshop on Hadron Structure at J-PARC'', Tsukuba, Ibaraki, Japan,
Nov. 30 - Dec. 2, 2005; Y.\ Shatunov, talk presented at the
``17th International Spin Physics Symposium (Spin 2006)'',
Kyoto, Japan, October 2-7, 2006. Plans for Drell-Yan measurements
in pion-nucleon scattering also exist for the COMPASS experiment.

\bibitem{BoeVog03}
  D.~Boer and W.~Vogelsang,
  Phys.\ Rev.\ D {\bf 69}, 094025 (2004).

\bibitem{Sterman}
see, for example, N.~Kidonakis, G.~Oderda and G.~Sterman,
  Nucl.\ Phys.\ B {\bf 525}, 299 (1998);
  Nucl.\ Phys.\ B {\bf 531}, 365 (1998).

\bibitem{mulders} C.~J.~Bomhof, P.~J.~Mulders and F.~Pijlman,
  Phys.\ Lett.\ B {\bf 596}, 277 (2004);
  Eur.\ Phys.\ J.\ C {\bf 47}, 147 (2006).

\bibitem{bacchetta}
  A.~Bacchetta, C.~J.~Bomhof, P.~J.~Mulders and F.~Pijlman,
  Phys.\ Rev.\ D {\bf 72}, 034030 (2005);
the $\cos(\phi_1^S)$ dependence in this paper must be replaced by a ${-}\cos(\phi_1^S)$ dependence everywhere.

\bibitem{bomhof} C.~J.~Bomhof and P.~J.~Mulders,
arXiv:hep-ph/0609206; to appear in JHEP.

\bibitem{Efremov}
  A.~V.~Efremov and O.~V.~Teryaev,
  Sov.\ J.\ Nucl.\ Phys.\  {\bf 36}, 140 (1982)
  [Yad.\ Fiz.\  {\bf 36}, 242 (1982)];
  A.~V.~Efremov and O.~V.~Teryaev,
  Phys.\ Lett.\ B {\bf 150}, 383 (1985).

\bibitem{qiu}
J.~Qiu and G.~Sterman,
Phys.\ Rev.\ Lett.\  {\bf 67}, 2264 (1991);
  Nucl.\ Phys.\ B {\bf 378}, 52 (1992);
Phys.\ Rev.\ D {\bf 59}, 014004 (1999).


\bibitem{twist3-06}
  C.~Kouvaris, J.~W.~Qiu, W.~Vogelsang and F.~Yuan,
Phys. Rev. D {\bf 74}, 114013 (2006).

\bibitem{VogYua05}
  W.~Vogelsang and F.~Yuan,
  Phys.\ Rev.\ D {\bf 72}, 054028 (2005); F.~Yuan, AIP Conf. Proc. 842, 383
  (2006).

\bibitem{boermulders} D.~Boer and P.~J.~Mulders,
  Phys.\ Rev.\ D {\bf 57}, 5780 (1998).

\bibitem{qvyprep} J.~W.~Qiu, W.~Vogelsang and F.~Yuan,
in preparation.

\bibitem{cteq5l}
  H.~L.~Lai {\it et al.}  [CTEQ Collaboration],
  Eur.\ Phys.\ J.\ C {\bf 12}, 375 (2000).


\end{thebibliography}
\end{document}